\documentclass[conference]{IEEEtran}
\IEEEoverridecommandlockouts

\usepackage{cite}
\usepackage{amsmath,amssymb,amsfonts}
\usepackage{graphicx}
\graphicspath{{figures/}}
\usepackage{booktabs}
\usepackage{stfloats} 
\usepackage{balance}
\usepackage{placeins}
\usepackage[hyphens]{url}
\usepackage[hidelinks,breaklinks=true]{hyperref} 

\def\BibTeX{{\rm B\kern-.05em{\sc i\kern-.025em b}\kern-.08em
    T\kern-.1667em\lower.7ex\hbox{E}\kern-.125emX}}

\begin{document}
\renewcommand\thesection{\arabic{section}}
\renewcommand\thesubsection{\thesection.\arabic{subsection}}

\makeatletter
\renewcommand{\@seccntformat}[1]{\csname the#1\endcsname\quad}
\makeatother

\setcounter{secnumdepth}{2}

\title{Hybrid Deep Learning-Federated Learning Powered Intrusion Detection System  for  IoT\slash 5G Advanced  Edge Computing Network}

\author{Rasil Baidar, Sasa Maric, Robert Abbas \\ rasil.baidar@live.vu.edu.au, s.maric@unsw.edu.au, robert.abbas@vu.edu.au}
\date{August 2025}

\maketitle

\begin{abstract}
The exponential expansion of IoT\slash 5G-Advanced applications has expanded the attack surface for DDoS, malware, and zero-day intrusions. We propose an intrusion detection system that fuses CNN, BiLSTM, and an autoencoder (AE) bottleneck within a privacy-preserving federated learning (FL) framework. The CNN–BiLSTM captures local and gated cross-feature interactions, while the AE emphasizes reconstruction-based anomaly sensitivity. Training occurs across edge devices without sharing raw data.

The fused model attains AUC \(=99.59\%\) and F1 \(=97.36\%\); confusion-matrix analysis shows balanced error rates with high precision and recall. Average inference time is \(\approx 0.0476\,\mathrm{ms/sample}\) on our test hardware—well within the \(<10\,\mathrm{ms}\) URLLC budget—supporting edge deployment. We discuss explainability, drift tolerance, and FL considerations for compliant, scalable 5G-Advanced IoT security.
\end{abstract}

\vspace{\baselineskip}
\vspace{\baselineskip}

\noindent\textbf{Abbreviations}
IoT — Internet of Things; DDoS — Distributed Denial of Service;
CNN — Convolutional Neural Network; LSTM — Long Short-Term Memory; BiLSTM — Bidirectional LSTM; AE — Autoencoder; FL — Federated Learning; GDPR — General Data Protection Regulation;
URLLC — Ultra-Reliable Low-Latency Communications; AUC — Area Under the ROC Curve;
F1-score — Harmonic mean of precision and recall.

\section{Introduction}
The exponential development and integration of the Internet of Things (IoT) has revolutionized modern industries and daily life with seamless interconnectivity between devices, systems, and services.

Low-latency communication, pervasive connection, and real-time data processing have all been transformed by the combination of 5G networks with the Internet of Things. Release 19 expands AI/ML-assisted RAN optimization, introduces low-complexity and low-power IoT devices such as advanced RedCap and ambient IoT, and further improves energy efficiency, spectral efficiency, and coverage optimization. 5G improves offering ultra-low latency, massive connectivity,
and better scalability for IoT devices. It also supports efficient spectrum use and
prioritizes traffic through network slicing. The integration of IoT ranges from smart homes to industrial automation, and its applications are growing at an unprecedented rate. The number of connected IoT devices is growing, with 18.8 billion by the end of 2024 and projected to reach 40 billion by 2030 as shown in recent forecasts~\cite{transformaIoTPage}. But this quick expansion has also made the attack surface larger, making it more susceptible to spoofing, botnets, Distributed Denial of Service (DDoS) attacks, and zero-day exploits. Intrusion attempts on IoT-enabled 5G networks have risen by more than 40\% yearly, according to the GSMA 5G Security Study (2024), underscoring the pressing need for more reliable and sophisticated IDS solutions. 
The centralized architectures used by traditional IDS techniques require sending raw data from IoT devices to a central server for processing. High latency, privacy issues, and scalability bottlenecks result from this. Furthermore, single-model deep learning methods could miss the varied and intricate patterns of contemporary cyberattacks. This study offers a hybrid deep learning model and proposes a federated learning architecture to improve intrusion detection in IoT\slash 5G environments in order to overcome these drawbacks. 

Many network security solutions do not have the ability to detect connected IoT devices or show which devices are communicating on the network. 
 IoT security challenges, including:
\begin{itemize}
    \item Weak authentication and authorization
    \item Lack of encryption
    \item Vulnerabilities in firmware and software
    \item Insecure communications
    \item Difficulty in patching and updating devices
\end{itemize}
 
~\cite{3GPP}IoT Reduced Capability is known as 5G RedCap (Reduced Capability), a technology designed for mid-tier 5G IoT use cases like industrial sensors and smart wearables by lowering device complexity and power consumption, enabling more scalable and efficient connectivity than full 5G devices but with higher capabilities than Massive IoT technologies.~\cite{3GPP} RedCap achieves this reduced capacity and complexity by limiting bandwidth to 20 MHz, eliminating carrier aggregation support, and using only one receive antenna instead of the two or four used in standard 5G devices.  
The key advantages of 5G RedCap:
\begin{itemize}
\item Enhanced 5G Ecosystem: It helps expand the 5G device ecosystem by enabling new types of connected devices.
\item Improved Battery Life: Lower complexity translates to better energy efficiency and longer battery life for connected devices. 
\item Scalable Connectivity: RedCap offers a scalable and efficient way to provide connectivity for a wide range of industrial and enterprise applications. 
\item Support for 5G Features: While simpler, RedCap still leverages advanced 5G network functionalities like enhanced positioning and network slicing for specific use cases. 
\end{itemize}
\subsection{Background and Literature Review}

The IoT expansion poses a significant security challenge. IoT is susceptible to threats such as generic threats, layers of architecture threats, and threats at the application layer~\cite{Iqbal}. As IoT networks continue its exponential growth, it is inevitable that the number of IoT cyber-attacks will rise as well. Intrusion attacks in IoT\slash 5G networks, including botnets, denial-of-service (DoS) attacks, and data breaches, are becoming more sophisticated, making traditional security measures like rule-based intrusion detection systems (IDS) and signature-based approaches ineffective.
paper~\cite{Gabriel } presents a deep learning
based model for network intrusion as well as a comparative
analysis of the performance of three major network intrusion
datasets using the proposed model. Results showed the model to
perform best for NSL-KDD, followed by UNSW-NB15 and CSECIC-
IDS2018 respectively. Model accuracy achieved for these
datasets were NSL-KDD (97.89\%), UNSW-NB15 (89.99\%), and
CSE-CIC-IDS2018 (76.47\%) was achieved.
Traditional intrusion detection systems (IDS) and conventional machine learning techniques often struggle to detect cyberattacks due to the dynamic and high-dimensional nature of IoT traffic. Signature-based detection systems, while being effective for known cyber threats, are generally inadequate for novel and evolving anomalies. Due to these limitations on current IDS, a need for more robust and adaptive anomaly detection methods capable of processing and analyzing complex IoT data in real-time is apparent.
This ~\cite{Syed} introduces an advanced deep hybrid learning model in
an asynchronous federated learning setup, aimed at improving the detection of cyberattacks
and ensuring robust data privacy.
Protecting the networks of tomorrow is set to be a challenging domain due to increasing cybersecurity threats and widening attack surfaces created by the Internet of Things and 5G, increased network heterogeneity, increased use of virtualisation technologies, and distributed architectures 
~\cite{Lam}.
~\cite{Kumar}
DeepTransIDS implements a Transformer-based Intrusion Detection System to analyse network traffic in 5G non-IP data delivery scenarios. Unlike traditional IDS approaches that rely on Convolutional Neural Networks, this work uses the self-attention mechanism of Transformers to enhance the classification performance for multi-class network intrusion detection.~\cite{gold} proposed a system that operated extremely well, attaining a 94\% detection precision, including those particular to IoT settings. The accuracy and dependability of ID are greatly improved by the suggested method's good handling of data imbalance, especially for uncommon or minor attack classes. The findings highlight how DL models can be used to protect IoT networks, which are frequently at risk from a number of attacks. The findings highlight how DL models can be used to protect IoT networks, which are frequently at risk from several attacks. Future research will concentrate on flexibility to novel attack types and real-time implementation. 
Deep learning (DL) has emerged as a promising approach for the detection of anomalies in IoT networks, offering high detection accuracy and adaptability. Integrated deep learning with IoT-security improves anomaly detection efficiency, leading to faster and more reliable identification of threats, has been highlighted in research published in Sensors~\cite{Rafique}. Similarly, a study in Scientific Reports found that deep learning-based anomaly detection models outperform traditional methods, especially in Industrial IoT (IIoT) environments~\cite{Zhen}.
The research in~\cite{Lambert} describes a federated learning system based on a Multi-Layer Perceptron with minimal computational complexity for detecting IoT botnet attacks. The proposed approach used an XGBoost model trained with repeated stratified k-fold cross-validation to choose optimal features for IoT botnet attack detection, followed by Principal Component Analysis to reduce dimensionality, hence reducing computational cost. 
~\cite{Satish} proposes a model that addresses the security issue of bot-related risks. 
Different machine learning algorithms, including K-Nearest Neighbour (KNN), Naive Bayes, and Multi-layer Perception Artificial Neural Network (MLP ANN), were utilized to create a model that was trained on the BoT-IoT dataset. 

~\cite{Mohammed} proposes a federated learning of routing protocol (Fed-RPL)-based gated recurrent unit (GRU) model for decentralized training rounds and a quantization method (Q-8bit) to reduce the number of weight updates, which can significantly reduce communication overhead while maintaining the local model with high accuracy. Meanwhile, the ensemble unit collects updates and chooses the best local model to improve global model accuracy. Our findings show that Fed-RPL outperforms traditional machine learning (ML) methods in privacy-preserving edge data, considerably lowers communication costs in non-IID contexts, and beats contemporary FL approaches in detection accuracy. 

Convolutional neural network (CNNs) excels at learning spatial features, long short-term memory (LSTM) networks capture temporal dependencies, and autoencoders (AE) are effective for unsupervised detection by learning compact representation of normal behavior. However, each of these approaches has its own strengths and limitations when applied in isolation.

To address the limitations of using single deep learning methods for anomaly detection, extensive research has been conducted on hybrid models that combine multiple deep learning methods. These hybrid models generally yielded better accuracy in detecting anomalies. For example, in~\cite{Hwang}, the author proposed a system, namely D-PACK, which combined CNN and AE, which outperformed prior studies and yielded near 100\% accuracy and less than 1\% FNR and FPR. Another study that combined LSTM and AE yielded 95.16\%, 99.12\%, and 75.13\% F1 score for the following datasets ECG5000, Wafer, Arrhythmia, respectively~\cite{Gao}.

~\cite{Shimin} propose
a personalized federated cross learning framework (pFedCross) for intrusion detection, to manage imbalanced
and heterogeneous data distributions. First, we present a collaborative model cross-aggregation algorithm for
personalized local model update, to solve the problem that one global model cannot always accommodate
all the incompatible convergence directions of local models.
Based on current research, we can conclude that a hybrid deep learning model is a feasible approach for anomaly detection in IoT. According to ~\cite{Vishwas}, relying entirely on methods using one model can result in high false positive rates and poor detection capabilities, leaving systems vulnerable to potential intrusions. A hybrid ensemble strategy combining Random Forest (RF) and Support Vector Machine (SVM) approaches shows promise for improving IDS efficacy. Despite the strengths of these two-model approaches, such as CNN-AE or LSTM-AE, current research lacks a fully integrated hybrid model that synergistically combines CNN, LSTM, and AE architectures. While CNN-AE~\cite{Hwang} and LSTM-AE~\cite{Gao} have been explored, there is still a gap in exploring a tri-model architecture that combines strengths of CNN, LSTM, and AE.

Additionally, several challenges persist in the application of deep learning methodology to IoT anomaly detection:

\begin{itemize}
    \item High computational costs associated with deep learning models, limiting their deployment on resource-constrained IoT devices.
    \item It is difficult to identify anomalies before they occur~\cite{Li}. As real\-world IoT traffic often has far fewer instances of attack data compared to normal behavior.
    \item Real\-time detection constraints, where deep learning models struggle to analyze data streams with low latency.
    \item Scalability concerns, as existing models often fail to generalize across diverse IoT environments and device types.
\end{itemize}

To address this gap, a hybrid deep learning model that integrates CNN, LSTM, and AE architectures to leverage their complementary strengths for more accurate anomaly detection in IoT networks. The key contributions of this research include:
Designing a novel model that combines CNN, LSTM, and AE for anomaly detection in IoT.
Evaluating the feasibility of the proposed hybrid model and comparison with other deep learning models.
Analyzing performance improvements in terms of accuracy and efficiency.
Conducting extensive evaluation of the proposed model using comprehensive set of datasets – UNSW\_NB15. By comparing and improving upon current deep learning models, this research aims to contribute to the development of lightweight, scalable, and an adaptive deep learning solutions for real-time anomaly detection with high accuracy and robustness.

IoT security solutions can be implemented by both device customers and manufacturers. 
The three types oF IoT security include:
\begin{itemize}
    \item Network Security: Users need to protect their devices against unauthorized access and potential exploitation. IoT network security implements a zero-trust security strategy to minimize the corporate attack surface.
\end{itemize}
\begin{itemize}
    \item Embedded: Nano agents provide on-device security for IoT devices. Runtime protection monitors the current state of the device and takes action based on anomalies to identify and remediate zero-day attacks.
\end{itemize}
\begin{itemize}
    \item Firmware Assessment: Firmware security starts with assessing the firmware of a protected IoT device. This finds potential vulnerabilities within an IoT device’s firmware.
\end{itemize}

\FloatBarrier
\section{System Model and Architecture}
As networks evolve toward 5G Advanced
and beyond, the diverse use cases—such as massive
Internet of Things deployments, ultra-reliable low-latency
communications, and enhanced mobile broadband—require
real-time decision-making and adaptive resource allocation.
Artificial intelligence (AI) provides a transformative solution
for optimizing network efficiency, performance, security, and
user experience and ensuring seamless operations across
complex, heterogeneous environments.
\begin{figure*}[!t]
    \centering
    \includegraphics[width=0.8\linewidth]{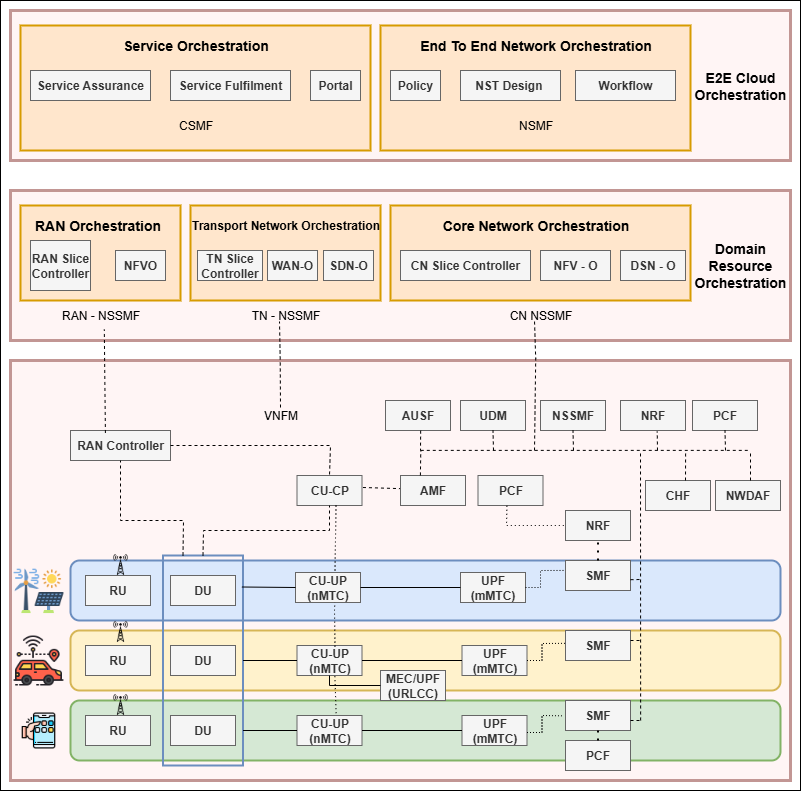}
    \caption{IoT-5G E2E Orchestration}
    \label{fig:Iot5GE2EOrechreation}
\end{figure*}

\subsection{IoT\slash 5G advanced Network Environment}

The combination of 5G and the Internet of Things (IoT) results in a robust network environment with expanded possibilities for a variety of applications. 5G's high speeds, low latency, and huge connectivity support a wide range of IoT devices and applications, allowing for real-time communication and data exchange. This confluence creates opportunity for smarter, more efficient, and environmentally friendly solutions across industries.
In the ever-evolving landscape of technology, IoT and 5G is converging and is entering into a new era of inter-connectivity and unprecedented transformative possibilities. The increasing intelligence, interconnection, and real-time capabilities of devices position the IoT-5G convergence to fundamentally alter everyday life, social interaction, and work practices.
The convergence of IoT and 5G is unlocking opportunities across healthcare, agriculture, smart cities, and manufacturing. by combining enhanced throughput, ultra-low latency, and massive device connectivity, 5G removes bottlenecks that previously constrained IoT deployments. This synergy enables reliable, real-time data exchange among devices, sensors, and applications, driving end-to-end automation, operational efficiency, and new classes of services.

As IoT and 5G converge, the resulting gains in throughput, latency, and device density unlock substantial opportunities across sectors; however, these benefits are accompanied by an expanded attack surface and heightened risks in data privacy, security, safety, and ethics. The rapid proliferation of heterogeneous endpoints and slice-based connectivity (eMBB, URLLC, mMTC) complicates authentication, key management, isolation, and policy enforcement at scale. This paper (i) delineates the application scenarios and performance gains enabled by IoT–5G integration, (ii) evaluates the operational and societal impacts of this convergence, and (iii) systematically analyzes emergent vulnerabilities and threat vectors, outlining mitigation strategies and design guidelines for resilient, privacy-preserving deployments.

\begin{figure*}[!b]
    \centering
    \includegraphics[width=0.8\textwidth]{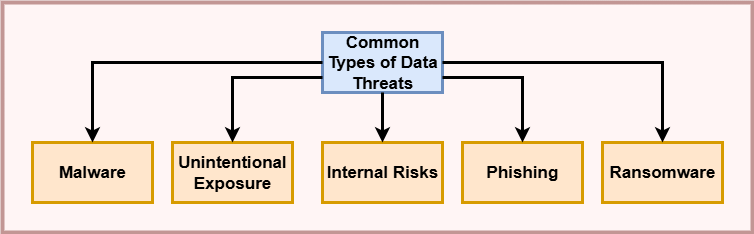}
    \caption{Common data threats.}
    \label{fig:typesofdatathreats}
\end{figure*}
The convergence of IoT and 5G is enabling impact at scale across healthcare, energy, manufacturing, and immersive media. By combining enhanced mobile broadband (eMBB), ultra-reliable low-latency communications (URLLC), and massive machine-type communications (mMTC), this ecosystem supports reliable, near-real-time data exchange among heterogeneous devices and applications. At the same time, the expansion of the attack surface and the societal stakes necessitate comprehensive security controls, clear regulatory frameworks, and sustained multi-stakeholder coordination.
High-throughput 5G (eMBB) allows devices and edge services to stream, analyze, and act on telemetry with minimal delay, improving operational efficiency and customer experience through faster monitoring, analytics, and closed-loop control. Low-latency 5G (URLLC) further enables mission-critical use cases—such as telesurgery support workflows, autonomous and cooperative mobility, and Industry 4.0 production lines—by reducing end-to-end response times and increasing reliability. Realizing these benefits in practice requires end-to-end assurance (e.g., network slicing, MEC placement, and deterministic QoS), coupled with rigorous security, safety, and compliance measures.
\begin{figure*}
    \centering
    \includegraphics[width=1\linewidth]{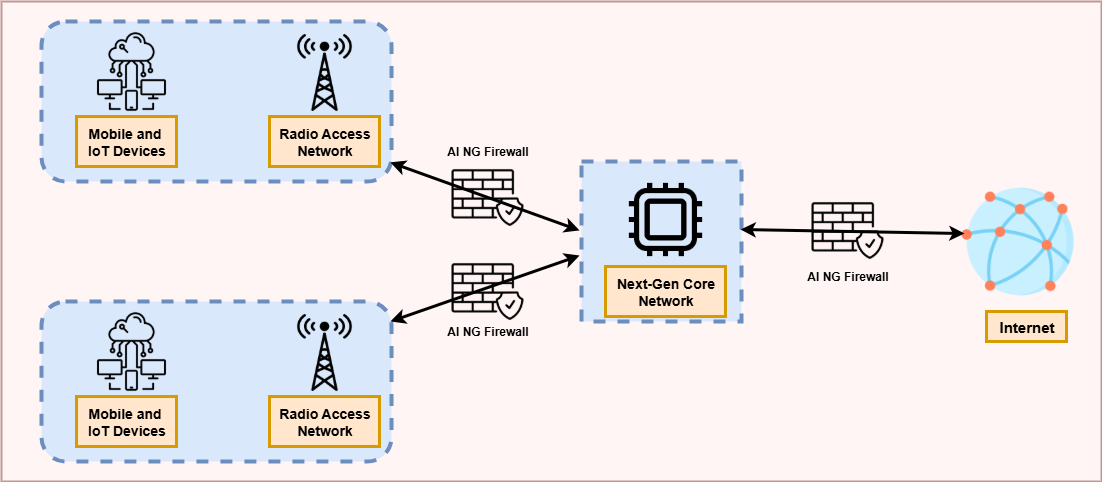}
    \caption{Common data transport links and threats}
    \label{fig:securingIOT}
\end{figure*}
The scalability of IoT in 5G-Advanced is a key advantage. Support for massive device density (mMTC) allows large-scale deployments of sensors, wearables, and embedded controllers without saturating the radio or core network. This capacity enables broader, higher-fidelity data collection and more accurate analytics and prediction.
These capabilities translate across domains. In healthcare, high-throughut and low-latency connectivity sustain telemedicine, remote diagnostics, and continuous patient monitoring. In smart cities, city-scale telemetry informs urban planning and resource allocation, improving energy efficiency and quality of life. In agriculture, logistics, and manufacturing, real-time monitoring and closed-loop control streamline operations, reduce costs, and raise productivity.

Realizing these benefits requires addressing a growing risk surface. The proliferation of endpoints and the volume of data intensify concerns around privacy, security, and compliance. Organizations should adopt defense-in-depth measures—strong device identity and attestation, end-to-end encryption, secure boot and update mechanisms, network slicing and micro-segmentation, least-privilege access, and data-protection practices (minimization, anonymization, and governance)—to ensure that the advantages of IoT in the 5G era outweigh the risks.

\subsection{NWDAF (Network Data Analytics Function) for IoT and 5G Advanced}
NWDAF (Network Data Analytics Function) for IoT and 5G Advanced provides AI/ML-powered data-driven intelligence for security and automation. 
3GPP-defined analytics function in the 5G core (Release 15 -18 Release, evolving further in 5G Advanced/6G).

Correlation: Analytics and prediction: Offers statistics, anomaly detection, trend forecasting, and QoS prediction for a variety of IoT\slash 5G applications.
Closed-loop automation: Provides insights for policy control, network slicing, and self-healing systems.
NWDAF in 5G Advanced Evolution

the edge and in the

Federated NWDAF: Uses federated learning to share insights without raw IoT data leakage.

Integration with AI/ML models: Supports hybrid ML (deep learning + graph ML) for IoT anomaly detection.

6G-ready enhancements: Extends to NTN (satellite IoT), ambient IoT, and AI-native networking.
NWDAF achieves this by getting the right data at the right cost to feed into the AI/ML \cite{transformaIoTPage}.

\subsection{IoT Security, Threat Model and Attack Scenarios}

The IoT Security solution integrates with next-generation firewalls to dynamically find and manage the IoT devices on your network. The IoT Security solution achieves high levels of accuracy using AI and machine-learning algorithms, even categorizing new IoT device kinds. And because it is dynamic, the IoT device inventory is constantly updated. IoT Security also automatically generates policy suggestions to restrict IoT device traffic and creates IoT device attributes for use in firewall policies. Accessing this solution requires an IoT Security membership. 

By nature, IoT devices are simple machines. As such, they lack sufficient
processing power to run on-board security. Most industry security
solutions require that a device run an agent or browser extension for
protection. IoT devices can’t run security agents and don’t support
browsers, making these solutions useless for IoT security. The default
passwords on IoT devices are seldom changed and updates are not
often installed. Most importantly, IoT devices typically broadcast their
IP addresses, making them an easy target for IP scans. With a zero-trust network and no visible IP address, IoT devices become invisible
to the outside world.

\begin{figure}
    \centering
    \includegraphics[width=0.9\linewidth]{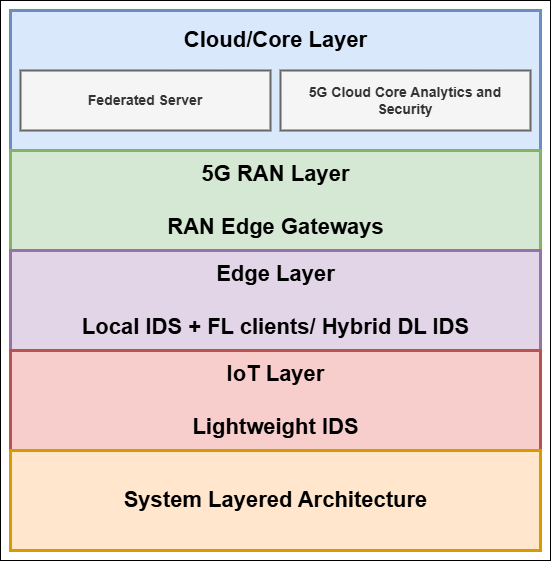}
    \caption{System Layered Architecture}
    \label{fig:cloudCoreLayer}
\end{figure}

For IT teams to effectively manage distributed IoT networks,
organizations need to implement data security practices that simplify
the setup and maintenance of IoT security solutions. With advanced
policies, organizations can securely connect third parties to remotely
manage devices. Zero-trust architecture is an ideal replacement
solution for VPNs~\cite{Sushil}.
One of the primary security concerns in the IoT 5G network is the potential compromise of sensitive data. As more devices collect and transmit data, the risk of data breaches becomes pronounced. Personal, financial, and proprietary business information is at stake, necessitating stringent encryption and data protection measures.

Furthermore, the speed and latency advantages of 5G can also expose devices to potential Distributed Denial of Service (DDoS) attacks. Attackers can leverage the high bandwidth of 5G to launch massive attacks, overwhelming devices and networks and disrupting services. This affects the availability of services and raises concerns about vulnerabilities in critical infrastructure.

In addition, IoT devices are notorious for their susceptibility to default or weak passwords, making them attractive targets for botnet attacks. Cybercriminals can compromise these devices and use them as part of larger botnets to launch attacks on other targets. The lack of proper security measures in the design and deployment of IoT devices exacerbates this risk.

As businesses embrace the benefits of the 5G IoT network, they must prioritize security measures. This includes implementing robust authentication protocols, regularly updating the device firmware, and performing comprehensive vulnerability assessments. In addition, organizations must ensure end-to-end encryption, secure device onboarding, and continuous network traffic monitoring for anomalies.
, IoT,

\section{Edge computing and Edge AI in IoT\slash 5G Networks}

Edge AI improves the Internet of Things (IoT) by moving artificial intelligence processing closer to the data source, allowing faster, more efficient, and secure operations. This involves placing AI models on edge devices, such as sensors and micro controllers, enabling them to analyze data and make decisions locally rather than relying entirely on cloud access. This~\cite{Daryll} paper studies the Distributed Denial of Service (DDoS) attack carried out over a 5G network and analyzes security attacks, particularly the DDoS attack. The aim is to implement a machine learning model capable of classifying different types of DDoS attacks and predicting the quality of 5G latency. The initial steps of implementation involved the synthetic addition of 5G parameters into the dataset. Subsequently, the data was label encoded, and minority classes were over sampled to match the other classes This strategy addresses critical IoT concerns such as latency, bandwidth, and security, resulting in increased performance and new opportunities for intelligent, autonomous IoT ecosystems.
Edge solutions enabled by 5G SDWAN were designed to extend private connection (5G and fixed) and cloud services to the customer and network edge, bringing computing, storage, network and marketplace services closer to clients. 
\begin{figure*}
    \centering
    \includegraphics[width=0.7\textwidth]{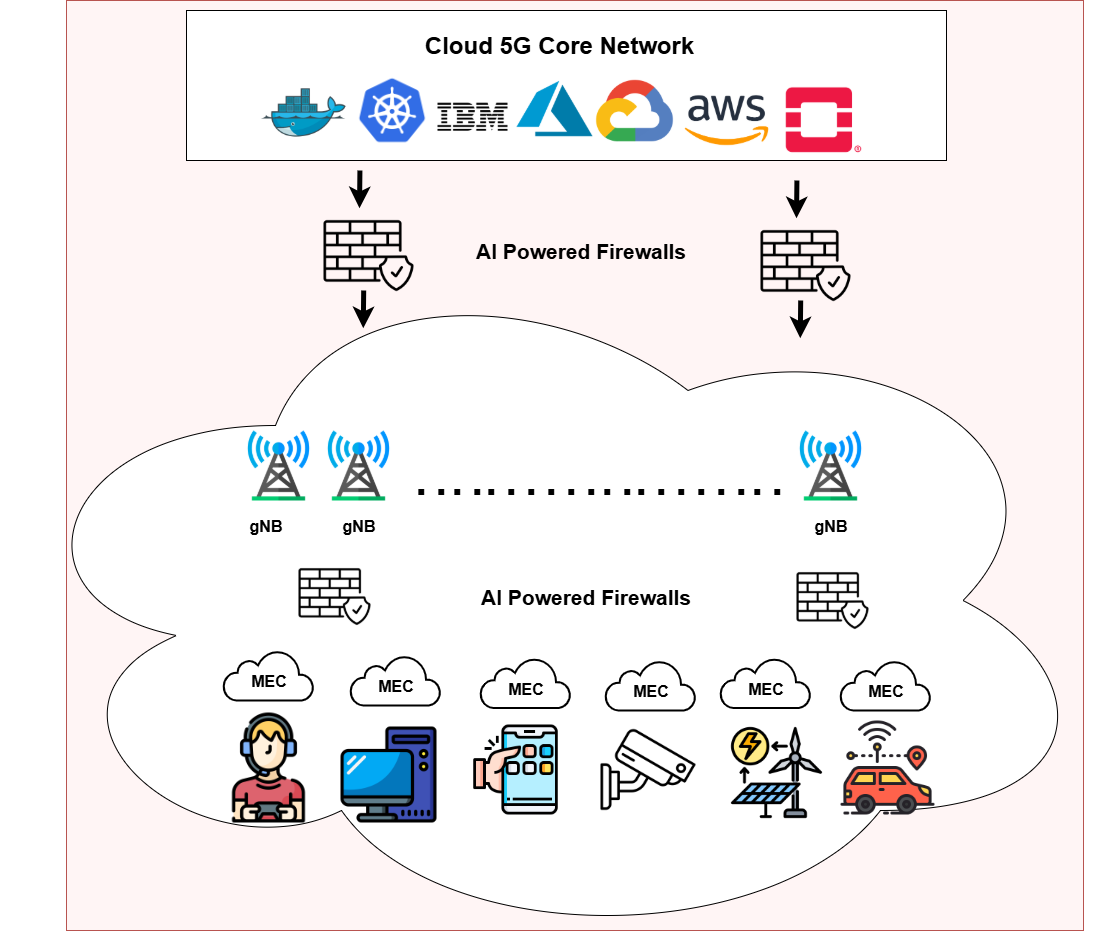}
    \caption{IoT\slash 5G advanced  edge services-enabled cloud Security }
    \label{fig:advanced_Edge_Services}
\end{figure*}
A 5G-based edge computing architecture diagram typically depicts user devices connecting to a local Radio Access Network (RAN) and then to a Multiaccess Edge Computing (MEC) (edge) server, which hosts applications and services closer to the user, before data is routed to the central 5G Core network and the cloud. This tiered structure, device tier, edge tier and central cloud tier, enables low latency processing by moving compute closer to data sources, complemented by the greater capacity of a 5G network and capabilities such as network cutting as in Figures ~\ref{fig:Iot5GE2EOrechreation}, ~\ref{fig:cloudCoreLayer}, and ~\ref{fig:IoT-5G_Edge_coputingcloud_security_architecture}.

\begin{figure*}[!t]
    \centering
    \includegraphics[width=0.9\linewidth]{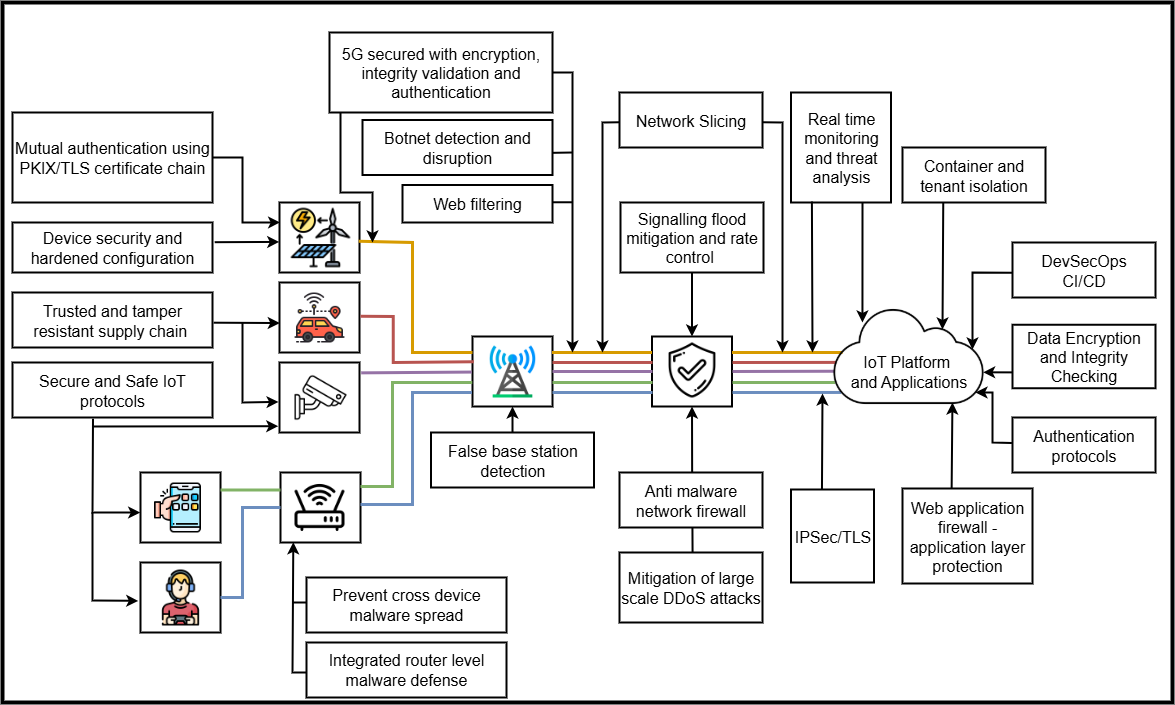}
    \caption{Security-vector-for IoT mobile networks application and platforms}
    \label{fig:securityVector}
\end{figure*}

IoT edge security is essential for protecting information and devices at the network's edge, where IoT devices are located. This covers threat detection, access control, and encryption. Lightweight and effective security solutions that support data encryption, sporadic connectivity, and physical vulnerability are required. With secure boot, static attestation, and certificate-based authentication, the Azure IoT Edge security architecture offers a strong security solution. For the highest security guarantees, it also connects with secure silicon hardware and provides an extension framework for further security services. 
\begin{figure*}
    \centering
    \includegraphics[width=1\textwidth]{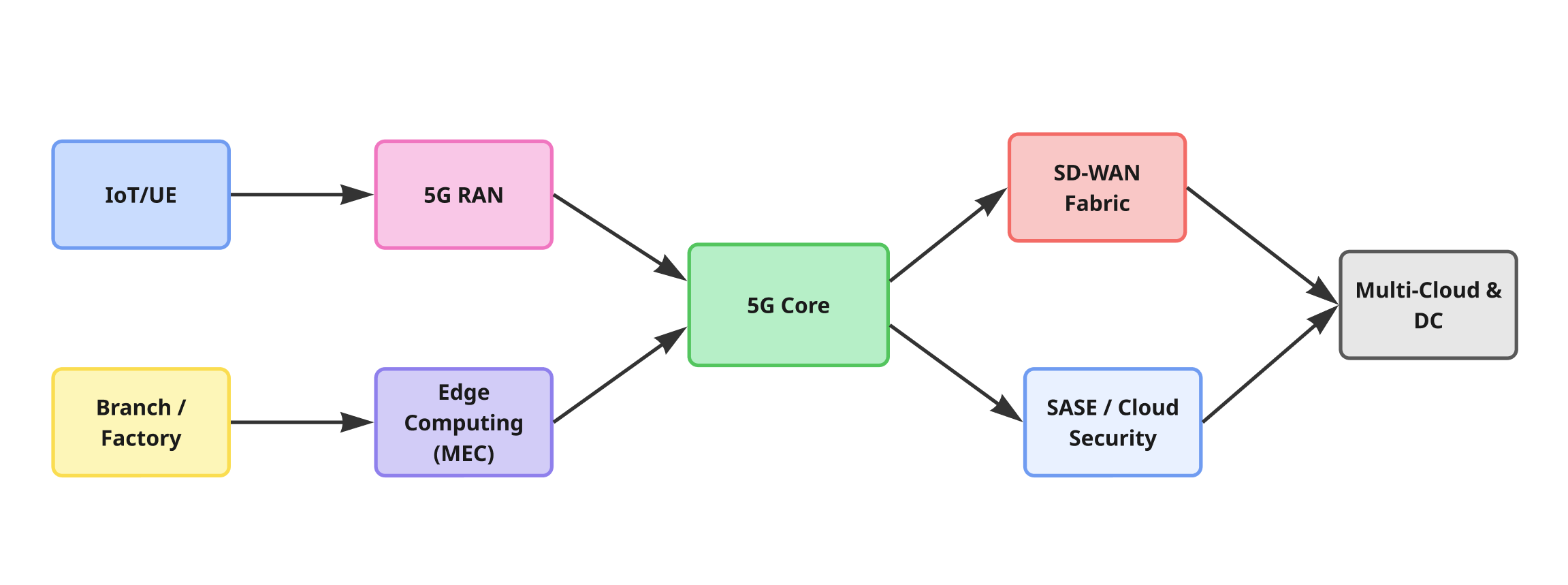}
    \caption{IoT\slash 5G edge computing cloud security architecture }
    \label{fig:IoT-5G_Edge_coputingcloud_security_architecture}
\end{figure*}

\FloatBarrier
\subsection{ \textbf{Edge computing AI in IoT and 5G}}
The interconnected and integrated system where AI, the Internet of Things (IoT), and edge computing are combined, creating a distributed platform for data processing and decision-making.
Edge AI improves IoT systems by enabling local real-time data processing, resulting in significantly lower latency, reduced bandwidth utilization, and enhanced privacy and security. This on-device intelligence also enhances reliability by enabling devices to operate autonomously during Internet disruptions, resulting in improved energy efficiency. Edge AI improves the intelligence, responsiveness, and resilience of IoT devices by processing data at its source.

 \begin{figure*}[!b]
     \centering
     \includegraphics[width=1\linewidth]{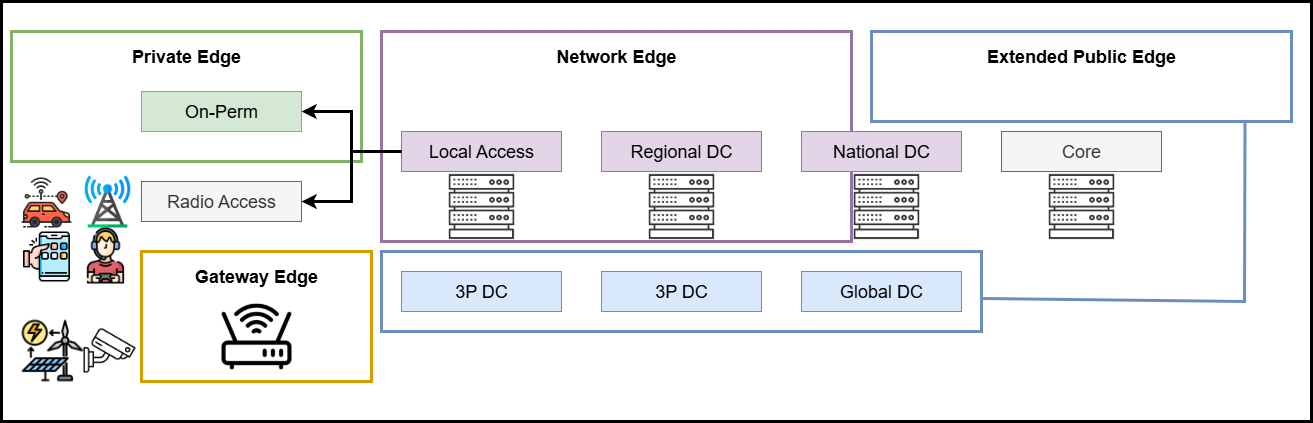}
     \caption{Four main locations for edge computing deployment}
     \label{fig:edgeComputing}
 \end{figure*}

\subsection{Federated Learning Framework}

An application framework and technique for federated learning facilitates distributed machine learning, enabling several clients to work together to train a common global model. These frameworks manage the process of training local models on various client data while protecting raw, sensitive information from being shared. Only model-related data is sent using the local data. updates (not data) to a central server for aggregation and distributing the updated global model back to the clients for improved performance while maintaining data privacy
An IoT federated learning framework is a decentralized system that enables AI models to be trained on multiple, distributed IoT devices without sharing their sensitive raw data, using local data to update a global model. Fig 9 shows the main locations to deploy edge computing physical infrastructure.

\subsection{Hybrid Deep Learning IDS Architecture}

Our proposed system, referred to as Precision AI, provides an advanced intrusion detection and prevention architecture designed for IoT and 5G network environments. At its core, the system integrates machine learning, hybrid deep learning, and generative AI to deliver real-time threat detection, prevention, and remediation across heterogeneous network, cloud, and endpoint infrastructures.

Incoming IoT and 5G traffic, typically sliced into multiple network partitions (eMBB, URLLC, and mMTC), is directed into the detection layer. Each slice presents unique security and quality-of-service requirements, demanding an adaptive and resilient defense strategy. The traffic is processed by a hybrid deep learning IDS architecuture
that fuses a 1D-CNN branch (capturing local cross-feature patterns), a BiLSTM branch (modeling temporal dynamics), and an autoencoder bottleneck trained with an auxiliary reconstruction loss to regularize the shared representation. This hybrid architecture enables precise classification of both known and unknown attack vectors. Furthermore, the training being primarily supervised, augmented by the self-supervised reconstruction objective to improve generalization and label efficiency. For edge deployment, the decoder is not included at inference and slice-aware thresholding is applied to satisfy differing false-alarm and latency budgets.

The architecture adheres to the principles of Zero Trust and Zero Touch security. Under the Zero Trust paradigm, every device and packet is continuously authenticated and authorized, eliminating reliance on perimeter-based defenses. Zero Touch operations are realized through the integration of Security Orchestration, Automation, and Response (SOAR) capabilities, which automate incident triage, mitigation, and recovery without requiring human intervention for common or repetitive threats. This automation significantly reduces the operational burden on security operations centers (SOCs) while improving response times.

A key innovation of the system is the integration of generative AI into both the training and operational pipelines. Generative models are used to synthesize realistic attack traffic, enabling the IDS to be trained against novel and adversarial threats that may not yet exist in the wild. During operations, generative AI also assists in summarizing threat intelligence, producing human-readable insights from large-scale telemetry and global threat feeds, and simplifying the user experience for analysts. This capability is augmented by explainable AI (XAI) mechanisms, which provide interpretable justifications for model outputs and support regulatory compliance as well as trust in AI-driven decision-making.

The operational workflow begins with ingestion of IoT\slash 5G network traffic, which is analyzed and classified by the hybrid deep learning IDS. Legitimate traffic is forwarded for application-level implementation, while malicious or anomalous flows are flagged. The system automatically categorizes such flows as denial-of-service (DoS), distributed denial-of-service (DDoS), connection dropping, or other IoT-specific attack classes. Depending on the classification, mitigation and prevention strategies are enacted in real-time. These include dynamic rate limiting, isolation of compromised nodes, traffic rerouting to avoid congestion or attack surfaces, or, in cases of highly suspicious or unverifiable traffic, black-holing, where traffic is discarded to prevent propagation of malicious payloads.

The system provides multi-layered defense coverage, addressing network-level attacks such as volumetric DDoS, signaling storms, and routing manipulation; cloud-level threats including API abuse and lateral movement; and endpoint-level threats such as malware injection and insider attacks. It is designed to handle advanced IoT botnets (e.g., Mirai and its variants), adversarial AI-based evasion attempts, and cross-slice 5G attacks that attempt to degrade service quality or disrupt mission-critical operations.

From an operational perspective, Precision AI integrates seamlessly with existing SIEM platforms (e.g., Splunk, ELK) and cloud-native monitoring systems (e.g., AWS GuardDuty, Azure Sentinel). It supports automated playbooks that trigger remediation workflows such as device patching, container isolation, or SDN-based traffic redirection. The orchestration of these processes is fully automated yet transparent, ensuring real-time situational awareness while reducing the cognitive and manual load on security personnel.
\begin{figure*}[!t]
        \centering
        \includegraphics[width=0.7\textwidth]{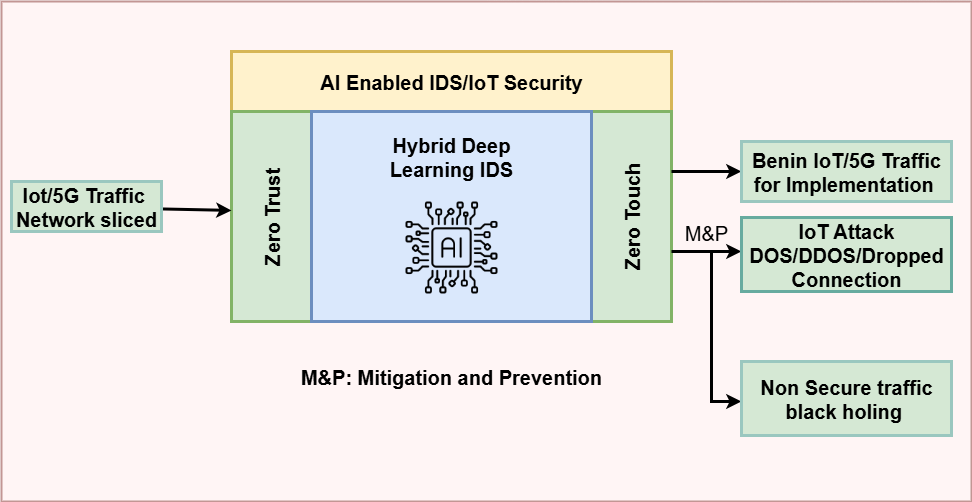}
        \caption{IDS operational work flow}
        \label{fig:IDSOperation}
\end{figure*}

The advantages of this architecture are manifold. Its scalability enables defense against threats targeting billions of IoT devices connected through dense 5G deployments. Its resilience is enhanced through self-healing network mechanisms, where compromised or congested paths are dynamically rerouted to maintain service continuity. Automation reduces false positives and alert fatigue in SOC environments, allowing human analysts to focus on high-level strategic defense. The system also embeds compliance-driven logging and audit trails to meet global regulatory requirements, including GDPR, HIPAA, and 5G security standards.

Despite these strengths, several research challenges remain. Ensuring robustness against adversarial AI attacks, where adversaries attempt to deceive the IDS with carefully crafted traffic, is a critical area of concern. Equally important is the need for privacy-preserving learning techniques, such as federated learning, to enable distributed IoT devices to collaboratively train detection models without exposing raw user data. Energy efficiency must also be considered, given the computational constraints of edge and IoT devices. Finally, secure cross-domain intelligence sharing mechanisms will be required to enable collaborative, global defense against rapidly evolving cyber threats, using approaches such as secure multi-party computation or homomorphic encryption.

\section {Proposed Methodology}
The best network security systems use AI and machine learning (ML) to analyze billions of daily events, improving defense against new attack trends. AI solutions can detect unusual patterns and actions that humans may miss. Machine learning enables quick action, discouraging adversaries and diverting their attention. This method reduces false alerts, allowing your team to prioritize serious dangers that require human involvement.

Using an IDS that is highly accurate for known attacks and capable of detecting new, unknown threats to address that. We utilize hybrid ML/DL techniques to develop a more intelligent, accurate, and robust system capable of addressing the complex challenges of the modern world, from securing 5G/ IoT networks to enabling autonomous driving.
An AI-powered  integrated platform can provide a unified configuration management aligned with best practices ensures uniform security across all settings, enable automatic intelligence sharing among technologies, and ease the policy-making process for intrusion prevention (IPS), using NGFW, intrusion detection, malware analysis, SaaS security
 , and IoT\slash 5G security.   

This research introduces an \textbf{Enhanced Intrusion Detection System (IDS)} which leverages a combination of three 3-different deep learning models (AE, CNN, BiLSTM) integrated within a \textbf{Federated Learning (FL)} framework for edge computing IoT / 5G~\cite{Ahmed}.
\subsection{Data Preprocessing and Feature Engineering}
For training our proposed hybrid model, we used the official UNSW-NB15 dataset,
particularly training and testing. The pre-processing of the dataset consisted of the following steps:
\begin{itemize}
    \item remove duplication and leakage
    \item mitigate numeric outliers
    \item control categorical cardinality
    \item encode and scale features consistently, and
    \item address class imbalance prior to model training
\end{itemize}
All the transformers are \emph{fit on the training partition only} and then applied to validation to avoid information leakage.

\paragraph{Schema and target.}
The binary intrusion label \texttt{label} is the target variable. Other non-predictive identifiers are excluded from the feature set. Specifically, we drop \texttt{id} and \texttt{attack\_col} which is a high-level descriptive column from the model inputs (the latter is retained only for subgroup reporting.)

\paragraph{De-duplication.}
Duplicate rows in the training split are removed. This prevents over-representation of repeated flows/records.
\paragraph{Outlier mitigation for numeric variables.}

\begin{figure*}
    \centering
    \includegraphics[width=0.9\linewidth]{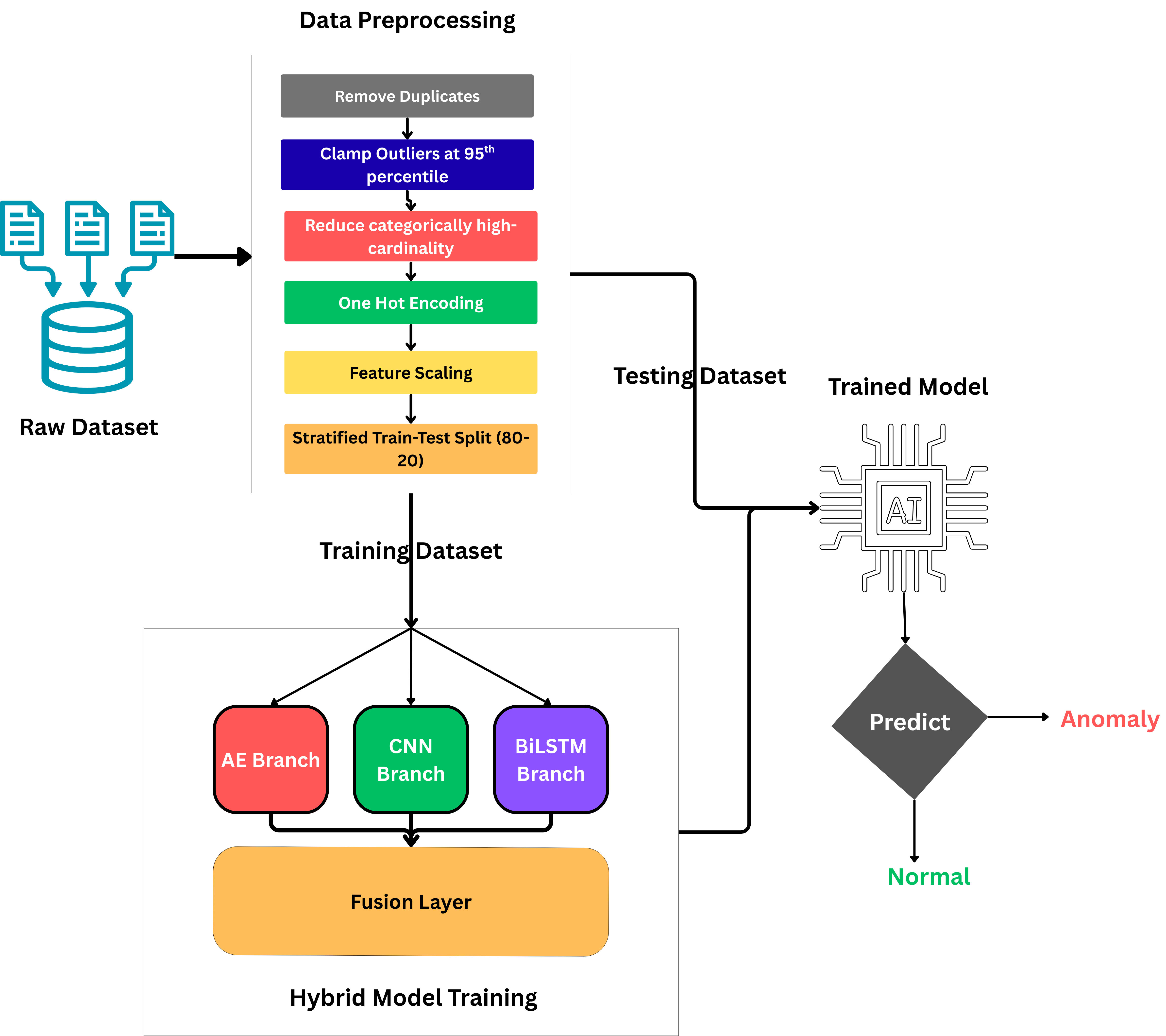}
    \caption{Proposed Hybrid Model}
    \label{fig:Tri-Hybrid_Model_Architecture}
\end{figure*}
In the UNSW-NB15 dataset, numeric features are often found to be heavy-tailed. To improve robustness, we apply a conditional winsorization: for a numeric feature x, if 
\[
\max(x) > 10\,\operatorname{median}(x) \;\land\; \max(x) > 10
\], values above the 95\textsuperscript{th} percentile are capped at that percentile. This preserves rank structure while limiting extreme influence.
\paragraph{Encoding and scaling.}
Categorical features are one-hot encoded with a reference category dropped (\texttt{drop="first"}). Numeric features are passed through unchanged at this stage. A \texttt{StandardScaler} is then fit on the training design matrix (after one-hot encoding) and applied to validation\slash test. This yields zero-mean, unit-variance inputs for all downstream branches of the hybrid model.
\paragraph{Train-test split and class imbalance.}
We perform an 80\slash 20 stratified split to maintain the attack/benign ratio across partitions. To correct for class imbalance in the training data, we apply SMOTE to the \emph{scaled} training set, generating synthetic minority class examples in the feature space. The test set is never oversampled or otherwise altered.

\subsection{Hybrid Deep Learning Model}
This research proposes an architecture consisting of \textbf{three parallel deep learning branches}- Autoencoder (AE), Convolutional Neural Network (CNN), and Bidirectional Long Short-Term Memory (BiLSTM) - all of which are converged through a \textbf{fusion mechanism} to form a robust intrusion detection system (IDS). The following section explains the role of each branch and how they work in cohesion.

\FloatBarrier
\subsubsection{Autoencoder (AE) Branch}
The AE branch comprises a dense encoder \verb|Dense(128) -> Dropout -> Bottleneck(64)|
and a decoder (\texttt{Dense(128) $\rightarrow$ Dense(53)} for reconstruction). The original high-dimensional input is compressed by the encoder into 64-dimensional latent embedding to captures the most salient traffic characteristics. By forcing reconstruction through this bottleneck, the AE learns to preserve the instrinsic structure of benign and malicious flows while discarding redundant or noisy attributes.
This branch contributes by:
\begin{itemize}
    \item Providing generalized embeddings that stabilize the input shape.
    \item Acts as a denoising filter, ensuring downstream branches operate on robust features.
    \item Enhances resilience to feature perturbations and facilitates detection of previously unseen intrusions.
\end{itemize}
This branch serves as the foundation of representation learning complementing the task-specific pattern recognition of CNN and BiLSTM.

\begin{figure}[b!]
    \centering
    \includegraphics[width=0.9\linewidth]{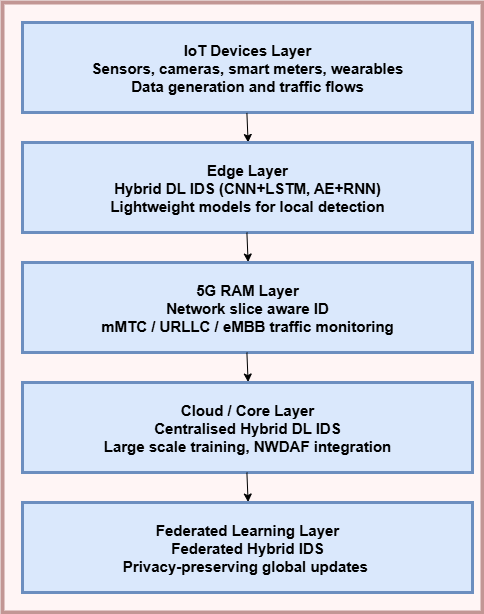}
    \caption{Hybrid Deep Learning architecture IDS for IoT edge 5G advanced networks}
    \label{fig:hybridDeeplearning}
\end{figure}

\subsection{CNN Branch}
In this branch, the high-dimensional input is reshaped into \verb|(53,1)| and is processed through stacked convolutional layers \texttt{Conv1D(64) ->\allowbreak MaxPooling ->\allowbreak Conv1D(128) ->\allowbreak GlobalMaxPooling ->\allowbreak Dense(64)}. CNN kernels act as a localized pattern detectors, uncovering correlation across subsets of network traffic features. These patterns are then abstracted by the pooling layers making the representations invariant to small input variations. This branch excels at capturing structural anomalies indicative of volumetric attacks such as Denial-of-Service (DoS) or Exploits.
This branch makes its contributions by:
\begin{itemize}
    \item Works synergistically with AE embeddings. Although AE offers a global view, CNN emphasizes fine-grained discriminative structures.
    \item Enhances the precision and recall of the model for attacks with different traffic signatures.
\end{itemize}

\subsection{ BiLSTM Branch}
The BiLSTM branch reshapes the input vector into $(1,53)$. It treats the features as a sequence of length one with 53 attributes. This is further processed by two stacked bidirectional LSTMs (\texttt{BiLSTM(64)} $\rightarrow$ \texttt{BiLSTM(32)}) followed by a dense layer (\texttt{Dense(64)}).
\newline Although the UNSW-NB15 dataset does not provide true temporal sequences, the BiLSTM branch interprets the feature vector as a structured sequence. The recurrent units apply bidirectional gating across the feature dimensions, producing enriched representations that capture dependencies beyond simple feedforward transformations. The second BiLSTM compresses these contextual embeddings into a 64-dimensional vector.
\newline The BiLSTM branch acts as a powerful transformation mechanism, introducing non-linear dependencies across features. While CNN highlights local spatial signatures, the BiLSTM emphasizes gated contextual interactions among features, making the overall model more sensitive to stealthy or low-profile intrusions. In conjunction with the AE and CNN branches, this ensures that both localized and distributed dependencies are represented in the fused decision space.

\subsection{Fusion Layer}
The latent embeddings produced by the three branches are concatenated: the Autoencoder bottleneck output (64 units), the CNN embedding (64 units), and the BiLSTM embedding (64 units). This results in a fused vector of 192 dimensions, which is passed through a fully connected integration block with L2 regularization and dropout (\texttt{Concatenate $\rightarrow$ Dense(128,L2) $\rightarrow$ Dropout(0.4) $\rightarrow$ Dense(1, sigmoid)}).

\subsection{Hybrid Deep Learning enabled IDS for IoT edge / 5G advanced networks} 
The combination of IoT edge computing and 5G Advanced networks enables unprecedented connectivity, low latency, and comprehensive device integration. However, these enhancements broaden the attack surface. Traditional IDS approaches struggle to handle multidimensional, heterogeneous, real-time traffic.  Hybrid Deep Learning (DL)-based IDS offers a scalable, adaptive, and intelligent approach to combating growing cyber threats in such scenarios.
where edge-cloud stability can make hybrid DL appropriate for distributed/federated deployments.

The success of AI for IoT network security and cloud security over the past five years has been attributed to the rapid development of AI-native network capabilities, the growth of data from the web, social media, and IoT sensors, as well as advancements in tensor-efficient computation with graphics processing units (GPUs) and a scalable, cloud-based AI toolchain.

\begin{figure*}[!t]
    \centering
    \includegraphics[width=\textwidth]{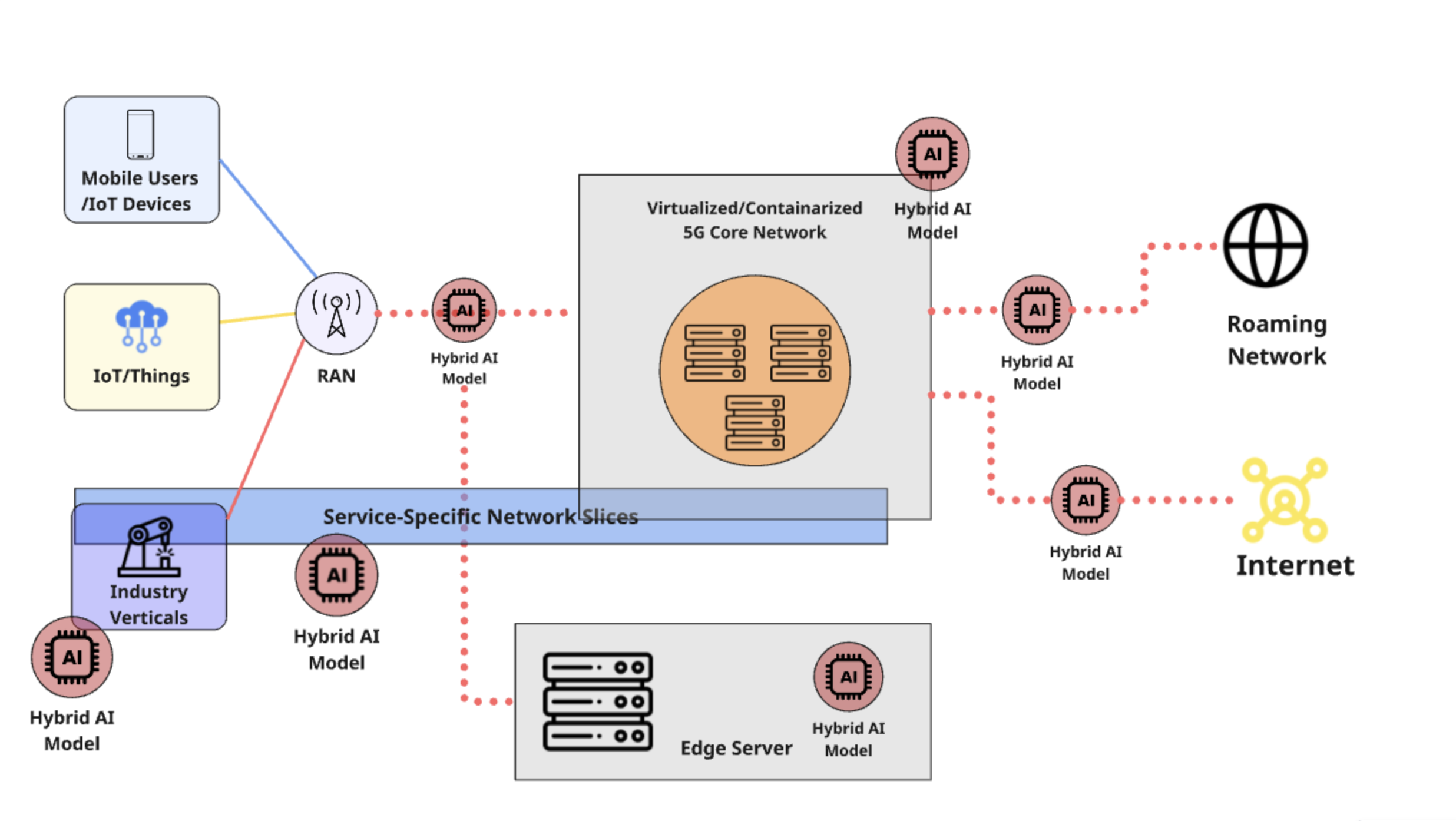}
    \caption{Hybrid DL implementation on IoT Edge 5G network architecture}
    \label{fig:iot-dl-implementation}
\end{figure*}

\FloatBarrier
\section{Results}
\subsection{Overall Performance}
Table ~\ref{tab:model-results} summarizes the performance of the proposed \emph{Fused AE-CNN-BiLSTM} model on the held-out test set. The model achieves an accuracy of 97.12\%, Precision of 98.45\%, Recall of 96.29\%, F1-score of 97.36\%, and AUC of 99.59\%. The near-perfect AUC indicates the model's excellent discriminative capability between benign and anomalous traffic.

\begin{table}
  \centering
  \small
  \setlength{\tabcolsep}{6pt}
  \begin{tabular*}{\columnwidth}{@{\extracolsep{\fill}}lc@{}}
    \toprule
    \textbf{Evaluation metric} & \textbf{Score (\%)} \\
    \midrule
    AUC       & 99.59 \\
    Precision & 98.45 \\
    Recall    & 96.29 \\
    F1        & 97.36 \\
    Accuracy  & 97.12 \\
    \bottomrule
  \end{tabular*}
  \caption{Performance of \emph{Fused AE--CNN--BiLSTM} model}
  \label{tab:model-results}
\end{table}
\subsection{Confusion Matrix and Error Breakdown}

\begin{figure}
    \centering
    \includegraphics[width=0.9\linewidth]{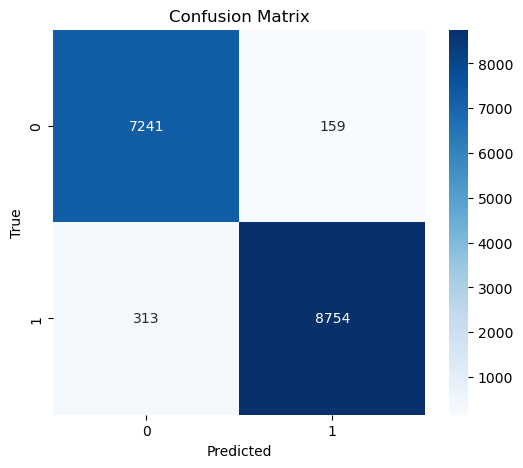}
    \caption{Confusion Matrix}
    \label{fig:confusion_matrix}
\end{figure}

Figure~\ref{fig:confusion_matrix} shows the confusion matrix with class~0 as benign and class~1 as attack. The counts are: TN=
7241, FP=159, FN=313, TP=8754 (total N=16,467). From these we obtain:
\begin{itemize}
  \item \textbf{True Positive Rate (TPR / Recall):} 
  \(\displaystyle \frac{8754}{8754 + 313} \approx 96.6\%\).
  \item \textbf{True Negative Rate (TNR):} 
  \(\displaystyle \frac{7241}{7241 + 159} \approx 97.9\%\).
  \item \textbf{False Positive Rate (FPR):} 
  \(\displaystyle \frac{159}{7241 + 159} \approx 2.1\%\).
  \item \textbf{False Negative Rate (FNR):} 
  \(\displaystyle \frac{313}{8754 + 313} \approx 3.5\%\).
\end{itemize}
The detector therefore errs slightly more on false negatives than false positives—a desirable trade-off can be achieved by moving the operating threshold (Section~\ref{sec:roc}).

\subsection{Learning Dynamics}
\begin{figure}
    \centering
    \includegraphics[width=1\linewidth]{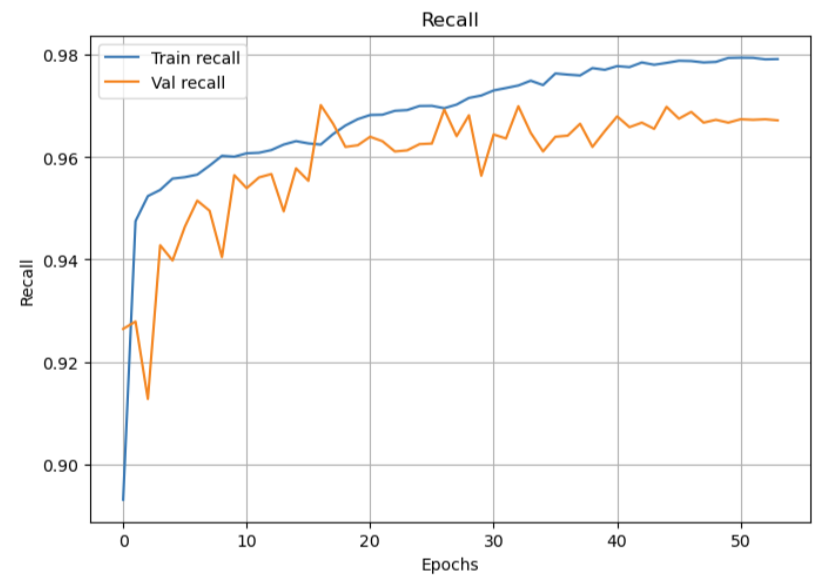}
    \caption{Recall Curve}
    \label{fig:recall}
\end{figure}

\begin{figure}
    \centering
    \includegraphics[width=1\linewidth]{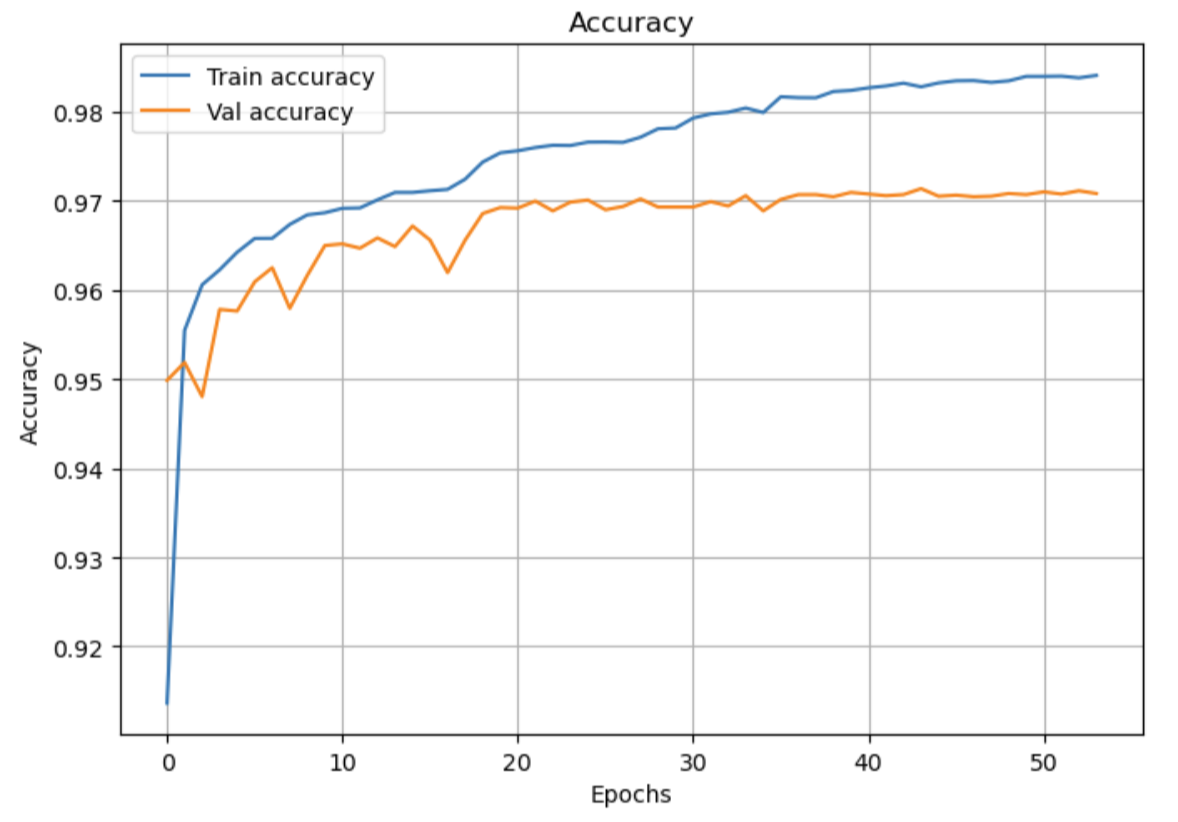}
    \caption{Accuracy Curve}
    \label{fig:acc}
\end{figure}

The training curves (Figures~\ref{fig:acc}-\ref{fig:recall}) show stable optimization without divergence. Validation accuracy plateaus around 30-35 epochs at \(\approx 97\%\), while validation recall converges near \(\approx 96.7\%\). The validation precision exhibits small oscillations early on, likely due to a class-mix\\threshold effects but stabilizes around \(\approx 98\%\). The generalization gap between the train and validation remains ~0.7-1.3 percentage points, suggesting minimal overfitting.

\begin{figure}
    \centering
    \includegraphics[width=1\linewidth]{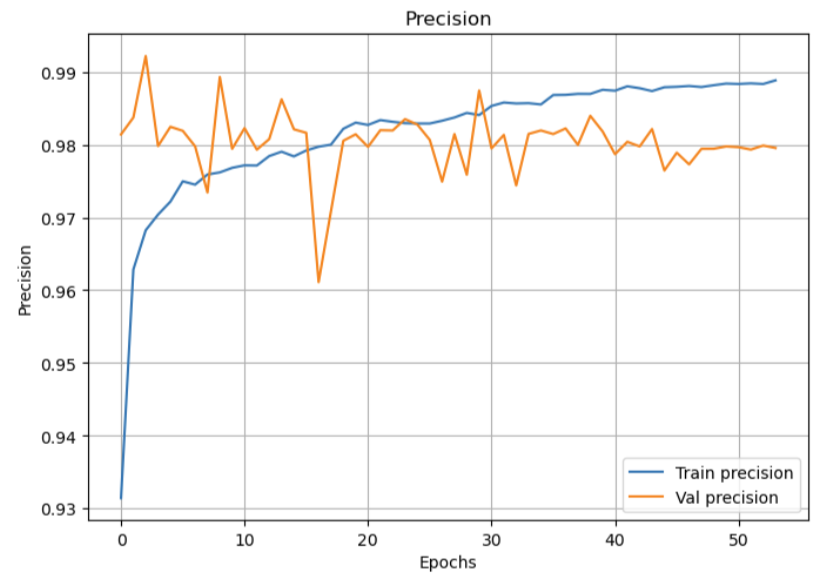}
    \caption{Precision Curve}
    \label{fig:precision}
\end{figure}

\subsection{ROC--AUC and Operating point}
\label{sec:roc}
\begin{figure}
    \centering
    \includegraphics[width=1\linewidth]{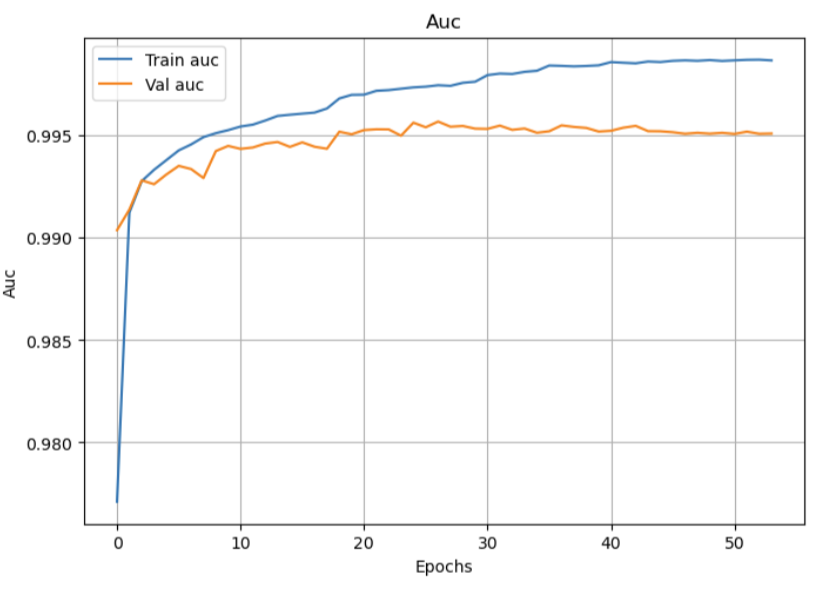}
    \caption{AUC-ROC Curve}
    \label{fig:auc-roc}
\end{figure}

The analysis of ROC (Figures~\ref{fig:auc-roc}) yields AUC = 0.9959, suggesting that the hybrid representation captures most discriminative structure in the traffic features. In deployment, the threshold for detection can be tuned to shift the FPR/FNR trade-off. For URLLC-critical scenarios favoring fewer false alarms, a slightly higher threshold reduces FPR at the cost of a modest recall drop; conversely, lowering the threshold maximizes recall for incident-response pipelines that can tolerate a small increase in alerts.

\subsection{Operational Throughput}
On the inference path, the model achieves an average processing time of \(\approx 0.0476\,\mathrm{ms/sample}\) on our test hardware. This is comfortably within the URLLC latency target of \(<\!10\,\mathrm{ms}\) per decision for 5G systems. This suggests that the model is suitable for edge deployment on gateway with modest compute budget.
\subsection{Interpretation}
Overall, the results show that:
\begin{enumerate}
  \item The hybrid design (AE bottleneck + CNN for local feature interactions + BiLSTM for temporal context) delivers strong separability (AUC~\(\approx 0.996\)) and balanced error rates (TPR~\(\approx 96.6\%\), TNR~\(\approx 97.9\%\)).
  \item The learning curves suggest the model converges smoothly and remains stable across epochs, with a small and acceptable generalisation gap.
  \item The combination of high precision (\(\approx 98.5\%\)) and high recall (\(\approx 96.3\%\)) yields an F1 of \(\approx 97.4\%\), reflecting reliable detection of both known and previously unseen attack behaviours.
  \item Inference latency meets edge constraints, supporting near–real-time filtering in 5G-Advanced IoT deployments.
\end{enumerate}

\paragraph{Notes for Reproducibility.}
All metrics are computed on the held-out test split. Confusion-matrix rates were derived from the counts in Fig.~\ref{fig:confusion_matrix}. The reported AUC is threshold-independent; accuracy, precision, recall, and F1 correspond to the operating threshold set via validation.

To support billions of low-cost, low-power, and passive IoT devices in large-scale scenarios, further research into ambient IoT and advancements in RedCap technologies are necessary. Human-centered and semantic service IoT-based.

\section{Conclusion}
To secure IoT/ 5G advanced cloud network from weaponized IoT device-initiated attacks and leverage deep visibility and granular control over cellular IoT traffic to discover and prevent attacks from known and unknown threats, command-and-control communications, denial-of-service attacks, and more.
This paper presented a hybrid deep learning intrusion detection system tailored for 5G-Advanced IoT edge environments. The model comprises of three complementary branches -- an Autoencoder (AE) for robust representation learning and reconstruction-based anomaly sensitivity, a Convolutional Neural Network (CNN) for local spatial pattern extraction, and a Bidirectional LSTM (BiLSTM) for gated cross-feature dependencies—followed by a compact fusion head. Additionally we outlined a federated learning (FL) training strategy to enable collaborative model improvement without sharing raw data, thereby enhancing privacy and scalability for distributed deployments.
On the held-out UNSW-NB15 test split, the fused model achieved \emph{AUC} = 99.59\%, \emph{Precision} = 98.45\%, \emph{Recall} = 96.29\%, \emph{F1} = 97.36\%, and \emph{Accuracy} = 97.12\%. The near perfect AUC score indicates that the model is highly discriminative against anomalous behaviors which is exactly what is needed for an IDS. The analysis of the confusion matrix indicates balanced error profiles with a slight bias toward minimizing false positives at the default operating point. Inference throughput averaged $\approx 0.0476\,\mathrm{ms/sample}$ on our test hardware, which is consistent with the stringent latency budgets typical of edge analytics in 5G-Advanced networks. Taken together, these results suggest that the hybrid design offers strong separability, competitive accuracy, and practical latency for gateway/MEC-class nodes.

In summary, the proposed AE–CNN–BiLSTM fusion, coupled with federated training, provides a practical path toward scalable, privacy-aware intrusion detection for 5G-Advanced IoT edge networks, combining strong detection performance with deployment-friendly latency.

\balance
\bibliographystyle{ieeetr}
\bibliography{ref.bib}

\begin{thebibliography}{10}

\bibitem{transformaIoTPage}
{Transforma Insights}, ``{Current IoT Forecast Highlights (IoT Connections Forecast 2023–2034)}.'' \url{https://transformainsights.com/research/forecast/highlights}, 2025.
\newblock Last updated: 01-Aug-2025; Accessed: Aug. 12, 2025.

\bibitem{3GPP}
{3GPP, RL 18,19,20}, ``{3GPP, Rl},'' {\em {3GPP}}, 2025.

\bibitem{Iqbal}
{Iqbal, Waseem Abbas, Haider Daneshmand, Mahmoud Rauf, Bilal Bangash, Yawar Abbas}, ``{An In-Depth Analysis of IoT Security Requirements, Challenges, and Their Countermeasures via Software-Defined Security},'' {\em {IEEE Internet of Things Journal}}, 2020.

\bibitem{Gabriel}
{Gabriel Chukwunonso Amaizu, Cosmas Ifeanyi Nwakanma, Jae-Min Lee, and Dong-Seong Kim}, ``{Investigating Network Intrusion Detection Datasets Using Machine Learning},'' {\em {IEEE Xplore}}, 2020.

\bibitem{Syed}
{Syed Muhammad Salman Bukhari a, Muhammad Hamza Zafar b, Mohamad Abou Houran c, Zakria Qadir d, Syed Kumayl Raza Moosavi b, Filippo Sanfilippo}, ``{Enhancing cybersecurity in Edge IIoT networks: An asynchronous federated learning approach with a deep hybrid detection model} [pdf] from journal homepage: www.elsevier.com/locate/iot,'' {\em {journal homepage: www.elsevier.com/locate/iot}}, 2024.

\bibitem{Lam}
{Jordan Lam, Robert Abbas}, ``{Machine learning based anomaly detection for 5g networks},'' {\em {arXiv preprint arXiv:2003.03474}}, 2020.

\bibitem{Kumar}
{Kumar Harshdeep a, Konatham Sumalatha b, Rohit Mathur }, ``{DeepTransIDS: Transformer-Based Deep learning Model for Detecting DDoS Attacks on 5G NIDD},'' {\em {Results in Engineering}}, 2025.

\bibitem{gold}
{Y. Gold Anand; S. Durai}, ``{Enhanced Security in IoT Networks through Optimized Deep Learning-Based Intrusion Detection} [pdf] from 2024 international conference on artificial intelligence and quantum computation,'' {\em {2024 International Conference on Artificial Intelligence and Quantum Computation}}, 2024.

\bibitem{Rafique}
{Rafique, Saida Hafsa; Abdallah, Amira; Musa, Nura Shifa; Murugan, Thangavel}, ``{Machine Learning and Deep Learning Techniques for Internet of Things Network Anomaly Detection—Current Research Trends},'' {\em {Sensors}}, 2024.

\bibitem{Zhen}
{Zhen, Chen; ZhenWan, Li; Jia, Huang; ShengZheng, Liu; HaiXia, Long}, ``{An effective method for anomaly detection in industrial Internet of Things using XGBoost and LSTM},'' {\em {Scientific Reports}}, p.~14, 2024.

\bibitem{Lambert}
{Lambert Kofi Gyan Danquah Iddrisu Danlard b a,* , Stanley Yaw Appiah , Emmanuel Kofi Akowuah a a a , Victoria Adzovi Mantey,Iddrisu Danlard b,Emmanuel Kofi Akowuah}, ``{Computationally Efficient Deep Federated Learning with Optimized Feature Selection for IoT Botnet Attack Detection} [pdf] from journal homepage: www.journals.elsevier.com/intelligent-systems-with-applications,'' {\em {/intelligent-systems-with-applications}}, 2025.

\bibitem{Satish}
{Satish Pokhrel, Robert Abbas, Bhulok Aryal}, ``{IoT security: botnet detection in IoT using machine learning},'' {\em {arXiv}}, 2021.

\bibitem{Mohammed}
{Mohammed Albishari a a , Mingchu Li a,b,* , Majid Ayoub,Ala Alsanabani Jiyu Tian}, ``{Federated deep learning models for detecting RPL attacks on large-scale hybrid IoT networks} [pdf] from journal homepage: www.elsevier.com/locate/comnet,'' {\em {Computer Networks Volume 254, December 2024, 110837}}, 2024.

\bibitem{Hwang}
{Hwang, Ren-Hung Peng, Min-Chun Huang, Chien-Wei Lin, Po-Ching Nguyen, Van-Linh}, ``{An Unsupervised Deep Learning Model for Early Network Traffic Anomaly Detection},'' {\em {IEEE Access}}, vol.~8, pp.~30387--30399, 2020.

\bibitem{Gao}
{Gao, Honghao Qiu, Binyang Barroso, Ramón J. Durán Hussain, Walayat Xu, Yueshen Wang, Xinheng}, ``{TSMAE: A Novel Anomaly Detection Approach for Internet of Things Time Series Data Using Memory-Augmented Autoencoder},'' {\em {IEEE Transactions on Network Science and Engineering}}, vol.~10, pp.~2978--2990, {2022}.

\bibitem{Shimin}
{Shimin Suna, Le Zhoua, Ze Wanga, Li Han}, ``{Robust intrusion detection based on personalized federated learning for IoT environment},'' {\em {computers-and-security}}, 2025.

\bibitem{Vishwas}
{Vishwas Sharma; Dharmesh J. Shahl}, ``{Novel Approach to Intrusion Detection Systems Using Hybrid Machine Learning Techniques} [pdf] from 2024 international conference on artificial intelligence and quantum computation,'' {\em {2024 International Conference on Artificial Intelligence and Quantum Computation}}, 2024.

\bibitem{Li}
{Li, Gen Jung, Jason J.}, ``{Deep learning for anomaly detection in multivariate time series: Approaches, applications, and challenges},'' {\em {Information Fusion}}, vol.~91, pp.~93--102, 2023.

\bibitem{Sushil}
{Sushil Shakya, Robert Abbas, Sasa Maric}, ``{A Novel Zero-Touch, Zero-Trust, AI/ML Enablement Framework for IoT Network Security},'' {\em {arXiv}}, 2025.

\bibitem{Daryll}
{Daryll Ralph D'Costa; Robert Abbasl}, ``{5G enabled Mobile Edge Computing security for Autonomous Vehicles} [pdf] from arxiv.org,'' {\em {arxiv.org}}, 2022.

\bibitem{Ahmed}
{Ahmed Asiri, Weiqi Wang , Feng Wu , Hiep Vo and Shui Yu}, ``{FedXAI for Detecting DDoS on IoT Edge Networks in Federated Learning},'' {\em {IEEE}}, 2024.

\end{thebibliography}

\end{document}